\documentclass[epj]{svjour}
\usepackage{graphics}
\usepackage{color}
\begin{document}
\title{Neutron Density Distributions of Neutron-Rich Nuclei Studied with the Isobaric Yield Ratio Difference}
\author{Chun-Wang MA, Xiao-Man BAI, Jiao YU, \and Hui-Ling WEI% etc
%\thanks{\emph{Present address:} Department of Physics, Henan Normal University, Xinxiang 453007 China}%
}                     % Do not remove
\offprints{C.-W. MA: machunwang@126.com}          % Insert a name or remove this line
\institute{Institute of Particle and Nuclear Physics, Henan Normal University, Xinxiang 453007 China
%    \and the second here
}
\date{Received: date / Revised version: \today}
\abstract{
The isobaric yield ratio difference (IBD) between two reactions of similar experimental setups
is found to be sensitive to nuclear density differences between projectiles. In this article,
the IBD probe is used to study the density variation in neutron-rich $^{48}$Ca. By
adjusting diffuseness in the neutron density distribution, three different neutron
density distributions of $^{48}$Ca are obtained. The yields of fragments in the 80$A$ MeV
$^{40, 48}$Ca + $^{12}$C reactions are calculated by using a modified statistical
abrasion-ablation model. It is found that the IBD results obtained
from the prefragments are sensitive to the density distribution of the projectile, while the IBD results from the final fragments are less sensitive
to the density distribution of the projectile.
 % end of PACS codes
} %end of abstract
\authorrunning{C.-W. MA \textit{et al.},}
\titlerunning{Neutron Density Distributions of Neutron-Rich Nuclei Studied with the Isobaric Yield Ratio ...}
\maketitle
\section{Introduction}
\label{intro}
Nuclear symmetry energy (NSE) is one of the hottest questions in the area of heavy-ion collisions
(HICs). The yield of a fragment is mainly determined by the free energy of a fragment, the chemical
potential properties of the source, and temperature in HICs above the intermediate energy \cite{MFM1,Huang-PRC11,PMar12PRCIsob,QMDFrg2011,Tsang07BET,MaCW12PRCT,MaCW13PRCT}.
In the many probes to determine the NSE, the isobaric yield ratio (IYR) \cite{Huang-PRC11,PMar12PRCIsob,Huang10,Huang10Powerlaw,MaCW11PRC06,NST13Lin,NST13WADA}
can provide cancelation for special energy terms in the free energy of a fragment. Huang \textit{et al.}
proposed the IYR method to determine the symmetry energy coefficients of the nearly symmetric fragments \cite{Huang10},
and this IYR method was extended to the neutron-rich fragment to understand the evolution of
symmetry-energy coefficient in neutron-rich nuclei \cite{MaCW12EPJA,MaCW12CPL06,MaCW13CPC}. The
IYR method is further developed, and improvements have been achieved to understand the IYR results \cite{RCIMa14,IBD13PRC,IBD13JPG,Souza12finite,Ma13finite1,Ma13finite2}.
The results of the NSE in Refs. \cite{PMar12PRCIsob,Huang10,PMar-IYR-sym13PRC,Mallik13-sym-IYR1,Mallik13-sym-IYR2}
show a large difference, but the NSE results determined by the isoscaling (IS) method (which uses
isotopic or isotonic ratios in HICs) and the isobaric yield ratio difference (IBD) method
are found to be similar \cite{IBD13PRC,IBD13JPG}. Besides, the IYR is also introduced to
determine the temperature of heavy fragments by using different approximations for the
free energy of fragments \cite{MaCW12PRCT,MaCW13PRCT,Ma2013NST}.

The IBD is found to be sensitive to nuclear density \cite{IBD13PRC,IBD13JPG,IBD14Ca}. The IBD, which
is defined as the differences between the IYRs in two reactions of similar experimental setups, is found
sensitive to nuclear density difference between projectiles. The IBD is defined as,
\begin{eqnarray}\label{IBDIS}
\Delta\mu_{21}/T&=\mbox{ln}[\frac{\sigma_{2}(A, I+2)}{\sigma_{2}(A, I)}]-\mbox{ln}[\frac{\sigma_{1}(A, I+2)}{\sigma_{1}(A, I)}], \nonumber\\
&=[\mu_{n2}-\mu_{n1}-(\mu_{p2}-\mu_{p1})]/T, \nonumber\\
&=(\Delta\mu_{n21}-\Delta\mu_{p21})/T, \hspace{1.3cm}
\end{eqnarray}
where $\mu_n$ ($\mu_p$) denotes the chemical potential of neutrons (protons), which depends on
both neutron (proton) density and temperature \cite{Bot02-iso-T,MBTsPRL01iso,rhoalpha}; $T$ is
the temperature; $\sigma(A, I)$ is the cross section of the fragment $(A, I)$ with $I (= N - Z)$.
Indexes 1 and 2 denote the reactions. $\Delta\mu_{21}/T$ is named as the IBD--$\Delta\mu_{21}/T$.
The probes based on $\mu_n$ ($\mu_p$) are found to be sensitive to NSE, for example, the isoscaling
parameters $\alpha\equiv\Delta\mu_{n21}/T$ and $\beta\equiv\Delta\mu_{p21}/T$ \cite{PMar12PRCIsob,Bot02-iso-T,Huang10NPA-Mscaling,HShanPRL,Iso-fluctuation13,ChenZQ10-iso-sym}.
In theory, $\Delta\mu_{21}/T = \alpha - \beta$, and this has been proved in fragments showing
the isoscaling phenomena \cite{IBD13PRC,IBD13JPG}. $\alpha$ ($\beta$) is
also correlated to the relative neutron (proton) density between reactions, which is
$e^{\alpha} = \hat{\rho}_n = \frac{ \rho_{n, 2}}{\rho_{n, 1}}$
($e^{\beta} = \hat{\rho}_p = \frac{\rho_{p, 2}}{\rho_{p, 1}}$). \cite{Bot02-iso-T,MBTsPRL01iso,rhoalpha}.
Equivalently, $\alpha = \mbox{ln} \rho_{n, 2} - \mbox{ln} \rho_{n, 1}$
($\beta = \mbox{ln} \rho_{p, 2} - \mbox{ln} \rho_{p, 1}$) denoting the difference between
the density of neutrons (protons).

A neutron-rich nucleus has a neutron-skin due to the large difference between its $\rho_n$ and
$\rho_p$ distributions. The neutron-skin thickness is also one of the important parameters to
study the NSE of the sub-saturation nuclear matter, but there are some difficulties in measuring
the neutron-skin thickness \cite{PVAHoroskin12,Roca11PVAskin,PrexPRL12,LW13skin,RZZ12skin,RZZ13skin,Tsang12skinsym,MaCW08CPB,DQF10PRC-nskin-SAA,MaCW10PRC,DAInskin14PRC}.
In this article, we study the sensitivity of the IBD probe to the neutron density variation
in $^{48}$Ca. In general, the quantum molecular dynamical (QMD) model \cite{QMD1,QMD2,QMD3,QMD4} (or its improved versions) and the antisymmetric
molecular dynamical (AMD) model \cite{AMD-ono,AMD-ono1,AMD-ono2,AMD-ono3,AMD-ono4}, in which the evolution of reaction can
be described, are usually adopted to simulate HICs. Most works using QMD concentrate on
the study of the phenomena involving light particles \cite{BALi08PR,QMD-appl11,QMD-appl12,QMD-appl13,QMD-appl14,QMD-appl21,QMD-appl22,QMD-appl23,QMD-appl24,QMD-appl25,QMD-appl31,QMD-appl32}.
For fragment with larger mass, QMD cannot well reproduce the isotopic or isobaric yield
distributions. Though AMD predicts the yields of fragments much better than QMD \cite{Huang10,AMD-mocko,NST13Lin,AMD-Roy1},
it needs a lot of time to gather enough events \cite{Huang10}. Besides the QMD and AMD models, the constrained molecular dynamics (CoMD) model may also be a good choice since CoMD cost less time than AMD, at the same time, CoMD can predict the yields of fragments well \cite{CoMDPRC011,CoMDPRC012,CoMDPRC013,CoMDPRC014,CoMD2013NST}. In the IBD analysis, the
yield of fragment with large mass is needed, which can not be easily fulfilled by
QMD and AMD. A modified statistical abrasion-ablation (SAA) model, which can well reproduce
the yield of fragment \cite{SAABrohm94,SAAGaim91,FangPRC00,Fang07JPG}, is adopted in this work. The
proton and neutron density distributions are distinguished in the SAA model \cite{FangPRC00,Fang07JPG,MaCW09PRC,MaCW09CPB},
which make it easy to change the density distribution. The SAA model will be briefly
described in Sec. \ref{modelSAA}. The IBD results are presented and discussed in Sec. \ref{results},
and a summary is presented in Sec. \ref{summary}.

\section{Model descriptions}
\label{modelSAA}

The SAA model has been described in Refs. \cite{SAABrohm94,SAAGaim91,FangPRC00}. To show
the results more clearly, we introduce the model briefly. The SAA model is a two-stage model
to describe the HICs above intermediate energies. The first stage is characterized by the
abrasion of nucleons in the projectile nucleus, and the hot prefragments are formed in this
stage. The second stage is characterized by the evaporation of light particles (n, p, $\alpha$),
in which the hot prefragment decay to final fragments. The projectile and target nuclei are
assumed to be composed of interacting tubes. Omitting the transverse motion of nucleons in the
tube, the collisions are described by independent interactions of tube pairs. Assuming the tube
to be infinitesimal, the average absorbed mass in a tube at a given impact parameter $\vec b$
is,
\begin{eqnarray}
<\Delta A(b)>=\int d^{2}s \rho_{n}^{P}(\vec s)[1-t_n(\vec s-\vec b)] \nonumber\\
+\int d^{2}s \rho_{p}^{P}(\vec s)[1-t_p(\vec s-\vec b)],
\end{eqnarray}
where $t_k(\vec s-\vec b)$ ($k = n, p$) is the transmission probability for neutrons (protons)
at $\vec b$, and is given by,
\begin{equation}\label{trans}
t_k(\vec s-\vec b)=\mbox{exp}\{-[\rho{_n^T}(\vec s-\vec
b)\sigma_{nk}+\rho{_p^T}(\vec s-\vec b)\sigma_{pk}]\},
\end{equation}
where $\rho^T$ is the nuclear-density distribution of the target integrated along the
beam direction, $\vec s$ and $\vec b$ are defined in the plane perpendicular to the beam,
and $\sigma_{k'k}$ are the nucleon-nucleon reaction cross sections. It is easy to see that
the yield of a prefragment is decided by the nuclear density distribution and $\sigma_{k'k}$.
The yield for a specific prefragment, which has $\Delta N$ of neutrons and $\Delta Z$
of protons removed from the projectile, is calculated as follows,
\begin{equation}\label{yieldisotope}
\sigma(\Delta N, \Delta Z)=\int d^2bP(\Delta N, b)P(\Delta Z,b),
\end{equation}
where $P(\Delta N, \mathit{b})$ [$P(\Delta Z, \mathit{b})$] is the probability
distribution for the abraded neutrons (protons) at a given $\mathit{b}$. These
probability distributions are a superposition of different binomial distributions
for all the tubes.

The Fermi-type density distribution is adopted for $\rho_n$ ($\rho_p$) \cite{Fermitype},
\begin{equation}\label{Fermi}
\rho_k(r)=\frac{\rho_k^0}{1+\mbox{exp}(\frac{r-C_k}{t_kf_k/4.4})},
~~k=n, p
\end{equation}
where $\rho_k^0$ is a normalization constant, $t_k$ is the diffuseness parameter,
and $C_k$ is the radius of the half density for the neutron (proton) density distribution.
$f_k$ is introduced to change the diffuseness and the neutron-skin thickness of
a nucleus \cite{MaCW08CPB,DQF10PRC-nskin-SAA,MaCW10PRC,DAInskin14PRC,MACW10JPG,MaCW11CPC}.
If $f_k = 1$, Eq. (\ref{Fermi}) is the original Fermi-type density distribution.

The evaporation of a prefragment is calculated by using a conventional statistical model
assuming the thermal equilibrium \cite{SAABrohm94,SAAGaim91}. The statistical hole-energy
model gives an average excitation energy of 13.3MeV per hole (due to abrasion). The excitation
energy of a prefragment is $E^{*}=13.3<A(b)>$MeV, with $<A(b)>$ being the average abraded
numbers of nucleons from the projectile. By using the Weisz\"{a}cker-Bethe mass formula \cite{Weiz-Bethe1,Weiz-Bethe2},
the binding energy of nuclei (A, Z), (A-1, Z-1), (A-1, Z) and (A-4, Z-2) are calculated, from
which the separation energy of neutron ($S_n$), proton ($S_p$) and $\alpha$ ($S_{\alpha}$)
of the nucleus (A, Z) can be decided. For a prefragment with $E^{*}$, the
most possible particle emitted is chosen according to min($S_n, S_p, S_{\alpha}$). The evaporation stops when the excitation energy
falls below the lowest particle threshold of the residual nucleus. This is an economical decay
method compared to the sequential decay method GEMINI \cite{gemini} or the statistical
multifragmentation model (SMM) \cite{SMM}. Similar methods have been adopted in works
dealing with the clusterization in the framework of the QMD model \cite{QMDSACA1,QMDSACA2,QMDSACA3}. After the
evaporation, the yield of the final fragment can be obtained. The SAA model can
well predict the yields of fragments in the 140$A$ MeV $^{40, 48}$Ca + $^{9}$Be and
$^{58, 64}$Ni + $^{9}$Be reactions \cite{MaCW09PRC,MaCW09CPB,WHL10}.

\section{Results and discussion}
\label{results}

\begin{figure}
\resizebox{0.5\textwidth}{!}{%
  \includegraphics{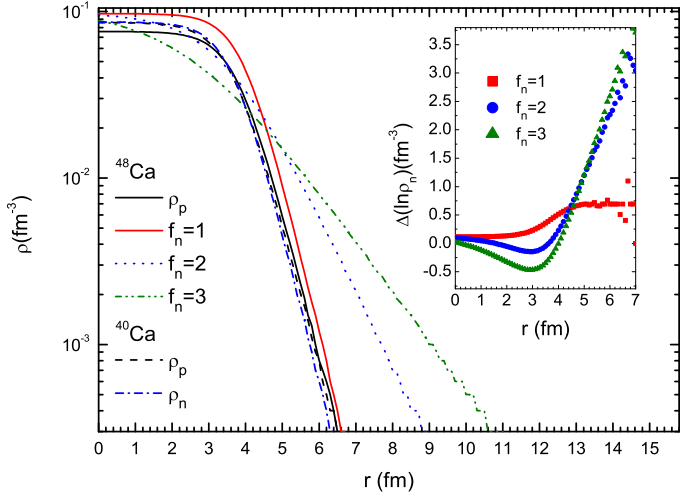}
}
%\vspace*{5cm}       % Give the correct figure height in cm
\caption{(Color online) The neutron and proton density distributions of $^{40}$Ca
and $^{48}$Ca according Eq. (\ref{Fermi}). For $^{48}$Ca, $f_n = $1, 2, and 3 are
plotted. In the inserted figure, the difference between the neutron density distributions
[$\Delta (\mbox{ln}\rho_n)$] of $^{40}$Ca and $^{48}$Ca are plotted, in which
$\Delta (\mbox{ln}\rho_n) = \mbox{ln}\rho_n(^{48}Ca)-\mbox{ln}\rho_n(^{40}Ca)$. $r$ denotes the
nuclear radius.}
\label{fndensity}       % Give a unique label
\end{figure}

The yields of fragments in the 80$A$ MeV $^{40, 48}$Ca + $^{12}$C reaction are calculated.
For $^{48}$Ca, $f_n =$ 1, 2 and 3 are adopted. The $^{40}$Ca reaction is labeled as
1, and the $^{48}$Ca reactions with a different $f_n$ are labeled as 2. Since
$\Delta\mu_{21}/T$ denotes the difference between the projectiles, we first discuss the
difference between $\rho_n$ ($\rho_p$) of $^{40}$Ca and $^{48}$Ca, which are plotted in
Fig. \ref{fndensity}. The $\rho_p$ distributions of $^{40}$Ca and $^{48}$Ca are very similar,
thus the difference between them will not be considered. For $^{48}$Ca, more neutrons are
pushed to the skirt region from the core with increasing $f_n$ resulting in the decrease
of $\rho_n$ in the core and making the neutron-skin thicker. In the inserted figure of
Fig. \ref{fndensity}, the results of $\Delta (\mbox{ln}\rho_n) = \mbox{ln}\rho_n(^{48}Ca)-\mbox{ln}\rho_n(^{40}Ca)$
are plotted. For $f_n = $1, $\Delta (\mbox{ln}\rho_n)$ is similar ($\sim$ 0) when r$<$3 fm,
but increases with $r$ in the range 3$ < r < $4.5 fm, and becomes similar again when $r >$ 4.5 fm.
For $f_n = $2 and 3, $\Delta (\mbox{ln}\rho_n)$ first decreases with the increasing $r$ when
$r < \sim$3.2 fm, then increases fast with $r$ almost linearly.

\begin{figure}
\resizebox{0.5\textwidth}{!}{%
  \includegraphics{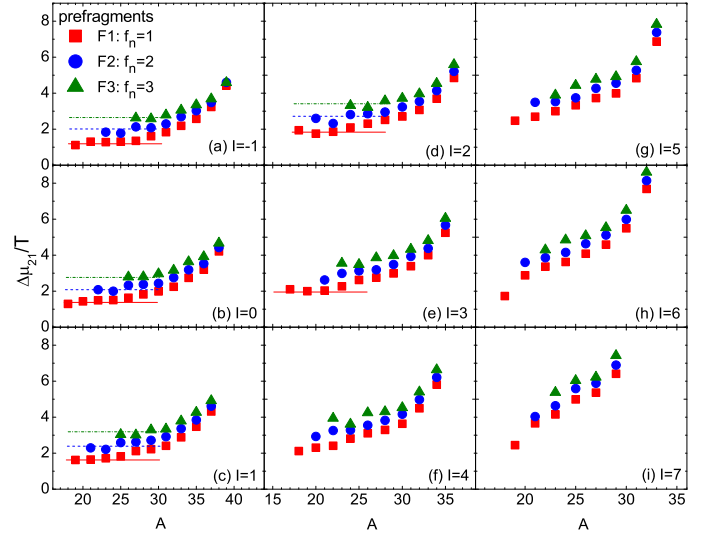}
}
%\vspace*{5cm}       % Give the correct figure height in cm
\caption{(Color online) $\Delta\mu_{21}/T$ determined from the prefragments produced
in the calculated 80$A$ MeV $^{48, 40}$Ca + $^{12}$C reactions. The squares, circles
and triangles denote the results for $f_n =$ 1, 2, and 3 in the $^{48}$Ca neutron
density distribution, respectively. From panel (a) to (i), the results are obtained
from the prefragments with $I$ from -1 to 7. The lines denote the plateaus in the $\Delta\mu_{21}/T$
distributions.}
\label{IBDpre}       % Give a unique label
\end{figure}

The yield of a prefragment is determined by the $\rho_n$ and $\rho_p$ distributions according
to Eq. (\ref{trans}), thus the probe constructed from prefragments is supposed to reflect
the density changes directly. It is concluded that $\Delta\mu_{21}/T$ is sensitive to the difference
between the $\rho_n$ of isotopic projectiles. The results of the IBD--$\Delta\mu_{21}/T$
are obtained by using the yields of prefragments and final fragments in the $^{48, 40}$Ca
reactions. For clarity, the $\Delta\mu_{21}/T$ for $^{48}$Ca with $f_n =$ 1, 2, and 3 are
labeled as F1, F2, and F3, respectively. The $\Delta\mu_{21}/T$ determined from the prefragments
are plotted in Fig. \ref{IBDpre}. When $I < 3$, the $\Delta\mu_{21}/T$ in the prefragment
with a relatively small $A$ form plateaus, while $\Delta\mu_{21}/T$ increases with $A$ in the
large $A$ fragments. When $I > 3$, the plateau of $\Delta\mu_{21}/T$ disappears, and
$\Delta\mu_{21}/T$ increases with $A$ almost linearly. The plateau in $\Delta\mu_{21}/T$ is
explained as the gentle change of $\rho_n$ in the cores of the projectiles, and the height of
the plateau in the $\Delta\mu_{21}/T$ distribution is assumed to reflect the difference between
$\rho_n$ of the projectiles \cite{IBD13PRC}. For the prefragments with $I$ from -1
to 7, the $\Delta\mu_{21}/T$ shows an obvious regular increase with $f_n$, which corresponds
to the enlarged $\Delta(\mbox{ln}\rho_n)$ between $^{48}$Ca and $^{40}$Ca. The plateaus in
$\Delta\mu_{21}/T$ of the $I =$ -1, 0, 1 and 2 prefragments also increase with $f_n$, but
the width of the plateau becomes shorter with the increased $f_n$. The enlarged $\Delta(\mbox{ln}\rho_n)$
can accounts for the increasing $\Delta\mu_{21}/T$ with $f_n$, and the increasing
$\Delta(\mbox{ln}\rho_n)$ in the $r < $3.2 fm explain the disappear of the plateau in
the neutron-rich fragments (see the inserted figure in Fig. \ref{fndensity}). Compared with the
heights of the IBD results ($\approx 2$) \cite{IBD13PRC} obtained from the measure fragments \cite{Mocko06}, the heights of the plateaus in
F1 are smaller, but those in F2 are similar.

\begin{figure}
\resizebox{0.5\textwidth}{!}{%
  \includegraphics{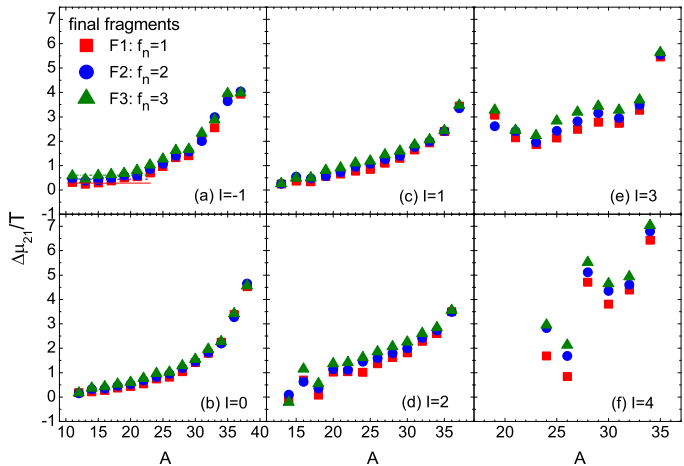}
}
%\vspace*{5cm}       % Give the correct figure height in cm
\caption{(Color online) The same as Fig. \ref{IBDpre} but for the results determined
from the final fragments. From panel (a) to (f), the results are from final fragments
with $I$ form -1 to 4. The lines in panel (a) denote the plateaus in the $\Delta\mu_{21}/T$
distributions.}
\label{IBDfinal}       % Give a unique label
\end{figure}

Though the prefragments can well indicate the density change in the projectile, they
are not measurable in the experiments. In SAA, the final fragments correspond to the measured
ones in experiment. The $\Delta\mu_{21}/T$ determined from the final fragments are
plotted in Fig. \ref{IBDfinal}. Most of the prefragments with $I >$ 5 can not survive
the evaporation process, thus only the $\Delta\mu_{21}/T$ obtained from the final
fragments with $I$ from -1 to 4 are plotted. In the final fragments, the distribution
of $\Delta\mu_{21}/T$ is also found increase regularly with $f_n$. The heights of the plateaus
are about 0.5, which are much smaller than the measured results in Ref. \cite{IBD13PRC}.
But the heights of the plateaus are larger than those in Ref. \cite{IBD14Ca},
which studied the density changes between calcium isotopes (the different $f_n$ makes a
larger $\Delta\rho_n$ compared with those in Ref. \cite{IBD14Ca}).

Now the phenomena that $\Delta\mu_{21}/T$ increases with $A$ in the large $A$ fragments will
be addressed. In this work, $f_p$ is not changed, thus in theory $\beta = 0$ and $\Delta\mu_{21}/T = \alpha$.
The underlying physics of the changing of $\alpha$ with $N$ should suit for that
of $\Delta\mu_{21}/T$. The values of $\alpha$ obtained form isotopic ratios of small $Z$ are
similar, but changes with $Z$ when $Z$ approximates to that of the projectile. This phenomena is
usually explained as that the large $Z$ fragments are produced in the peripheral reactions,
which correspond to the surface region of the projectile where $\Delta(\mbox{ln}\rho_n)$ changes
quickly. We have also explained the increasing $\Delta\mu/T$ (or $\alpha-\beta$) with $A$ in a
similar way by consider the central and peripheral collisions separately \cite{IBD13PRC,IBD13JPG,IBD14Ca}.
The fragments with small $A$ are mainly produced in the central collisions which is almost
not influenced by the neutron-skin, while the fragments having $A$ near the projectile are
governed by the neutron-skin \cite{Ma13finite1,Ma13finite2,MaCW10PRC,MACW10JPG}. The neutron-skin is also
found to make the IYR increase nonlinearly \cite{Ma13finite1,Ma13finite2}.

At last, we discuss the evaporation effects in $\Delta\mu_{21}/T$. The secondary decay
influences the isoscaling result greatly. The isobaric yield distributions and the
resultant IYRs can also be largely modified by the decay process \cite{MaCW13CPC,IBD14Ca}.
Though the IYRs are not compared directly in this work, it is easy to find the mass ranges
of the final fragments are larger than those of the prefragments. The $\Delta\mu_{21}/T$
obtained from the final fragments are much smaller than those from the prefragments, and are
sensitive to the decrease of $f_n$, which has been found in the $\Delta\mu_{21}/T$ obtained in the
reactions induced by the calcium isotopes \cite{IBD14Ca}. In the decay process,
the binding energy of nucleus is used by omitting its temperature dependence. In fact, the
binding energy of a light nucleus is influenced much more easily by temperature than that of a heavy
nucleus \cite{MaCW12PRCT}, which potentially influence the quality of the SAA model. For
fragments in a specific $Z$-chain produced in the 140$A$ MeV $^{40, 48}$Ca ($^{58, 64}$Ni) + $^{9}$Be
reactions \cite{AMD-mocko,Mocko06}, the yields of the small-$A$ isotopes have relative large
errors compared with those of the large-$A$ isotopes \cite{WHL10}. Thus the evaporation in
the present SAA model make it difficult to conclude wether the IBD probe is useful in detecting
the neutron density in a neutron-rich nucleus. Further investigation of the evaporation
effects in the IBD probe is suggested.

\section{summary}
\label{summary}

In summary, the IBD probe is used to study the density change in the neutron-rich $^{48}$Ca
nucleus. The SAA model is used to calculate the yields of fragments in the 80$A$ MeV $^{40, 48}$Ca + $^{12}$C
reactions. $f_n =$ 1, 2 and 3 are adopted to adjust the neutron density distributions in
$^{48}$Ca. The results of IBD--$\Delta\mu_{21}/T$ are obtained by analyzing the yields of the
prefragments and final fragments produced in the $^{40}$Ca and $^{48}$Ca reactions. The
$\Delta\mu_{21}/T$ distribution is explained by using the $\Delta(\mbox{ln}\rho_n)$ between
$^{48}$Ca and $^{40}$Ca, in which the plateau of $\Delta\mu_{21}/T$ is explained as the gentle
change of $\Delta(\mbox{ln}\rho_n)$ in the core of the projectiles, and the increasing
$\Delta\mu_{21}/T$ with $A$ is explained as the enlarged $\Delta(\mbox{ln}\rho_n)$ between the
surface region of the projectiles. The decay process influences the $\Delta\mu_{21}/T$ greatly.
The $\Delta\mu_{21}/T$ obtained from prefragments is sensitive to $f_n$, while the $\Delta\mu_{21}/T$
obtained from the final fragments is less sensitive to $f_n$. In the present status of the
SAA model, it is not easy to conclude whether the IBD probe can be used to detect the density
change in a neutron-rich nucleus. In addition, the improvement of evaporation in the SAA model
is suggested.

\section*{Acknowledgments}
This work is supported by the Program for Science \& Technology Innovation Talents in
Universities of Henan Province (13HASTIT046), and the Young Teacher Project in Henan
Normal University.


\end{document}